\documentclass[useAMS,usenatbib]{mn2e}
\usepackage{epsfig}

\title[Simulating seed BH formation: an improved chemical model]{Simulating the formation of massive seed black holes in the early Universe. I: An improved chemical model}
\author[S.~C.~O. Glover]{Simon C. O. Glover \\
Universit\"{a}t Heidelberg, Zentrum f\"{u}r Astronomie, Institut f\"{u}r Theoretische Astrophysik, \\ Albert-Ueberle-Stra{\ss}e 2, 
69120 Heidelberg, Germany
}

\begin{document}
\maketitle

\begin{abstract}
The direct collapse model for the formation of massive seed black holes in the early Universe
 attempts to explain the observed number density of supermassive black holes (SMBHs) at $z \sim 6$
 by assuming that they grow from seeds with masses $M > 10^{4} \: {\rm M_{\odot}}$ that 
 form by the direct collapse of metal-free gas in atomic cooling halos in which H$_{2}$ cooling 
 is suppressed by a strong extragalactic radiation field. The viability of this model depends on
 the strength of the radiation field required to suppress H$_{2}$ cooling, $J_{\rm crit}$: if this
 is too large, then too few seeds will form to explain the observed number density of SMBHs.
 In order to determine $J_{\rm crit}$ reliably, we need to be able to accurately model the
 formation and destruction of H$_{2}$ in gas illuminated by an extremely strong radiation 
 field. In this paper, we use a reaction-based reduction technique to analyze the chemistry of
 H$_{2}$ in these conditions, allowing us to identify the key chemical reactions that are 
 responsible for determining the value of $J_{\rm crit}$. We construct a reduced network of 
 26 reactions that allows us to determine $J_{\rm crit}$ accurately, and compare it with previous
 treatments in the literature. We show that previous studies have often omitted one or more
 important chemical reactions, and that these omissions introduce an uncertainty of up to
 a factor of three into previous determinations of $J_{\rm crit}$.
\end{abstract}


\section{Introduction}
In recent years, infrared sky surveys such as UKIDSS have discovered
a number of quasars at redshifts $z > 6$, including the current
record holder, a bright quasar at $z = 7.085$ \citep{mort11}. 
The presence of these objects indicates that supermassive black holes (SMBHs)
with masses of order $10^{9} \: {\rm M_{\odot}}$ were already present in
the high redshift Universe at a time when the age of the Universe was only
$\sim 800$ million years. However, the existence of these objects is difficult to 
understand within the standard $\Lambda$CDM cosmological model if one 
assumes that they grew from initial seeds with masses comparable to that of the 
Sun. Models for accretion onto black holes show that the characteristic growth
time for a black hole accreting at the Eddington limit is unlikely to be much smaller
than $\sim 50$ million years, unless one assumes an unreasonably low value for
the radiative efficiency of the accretion process. This means that a seed black 
hole formed at very high redshift can grow by at most a factor of $\sim 10^{7}$ by
the time that the high redshift SMBHs are observed. To produce SMBHs
with the observed masses, we therefore require seed black holes with masses
$M \sim 100 \: {\rm M_{\odot}}$ or higher \citep{th09}. In addition, if one accounts for the 
fact that radiative and mechanical feedback from the progenitor stars of these seed black 
holes will remove much of the gas from their local environment, it becomes difficult 
to see how the required high accretion rates can be sustained at high redshift, further 
increasing the required seed mass \citep[see e.g.][]{awa09}.

This problem can be avoided if we assume that instead of forming as the remnants
of Pop.\ III stars, the seed black holes that later grow into SMBHs form directly
from the monolithic collapse of primordial gas. For this so-called direct collapse
model to work, however, it is necessary that the collapsing gas remain warm
throughout the collapse, with a temperature of 5000--10000~K \citep{bl03}. This
keeps the Jeans mass high, preventing the gas from fragmenting into stellar mass
clumps, and leading to the formation of a central seed black hole with a mass of
order $10^{4} \: {\rm M_{\odot}}$ or more \citep{begel06,choi13,latif13a}. 
For this model to work, we therefore
require a minihalo with a virial temperature in excess of $10^{4} \: {\rm K}$ (so that
the gas does not simply become thermally supported at low densities) that has
not been enriched with heavy elements, which would otherwise provide efficient
cooling down to temperatures $T \ll 10^{4} \: {\rm K}$ \citep{om08}. In addition, it is
necessary that the formation of H$_{2}$ in the gas be strongly suppressed, as otherwise
H$_2$ cooling alone would be sufficient to cool the gas down to temperatures far
below $10^{4} \: {\rm K}$ \citep{om01,bl03}. The most viable mechanism known for
producing the required suppression of H$_{2}$ formation invokes the presence of a very strong
extragalactic UV field that immediately photodissociates almost all of the H$_{2}$
molecules forming in the gas \citep{bl03,visbal14}, and that also suppresses the formation of
H$_{2}$ by photodissociating the intermediate H$^{-}$ and H$_{2}^{+}$ ions \citep{sbh10}.

Numerical simulations (see e.g.\ \citealt{bl03,rh09,sbh10,latif13a,latif13b,bec14,regan14}, as well as the recent reviews
by \citealt{volo10}, \citealt{zh13} and \citealt{greif14}) have confirmed many of the details of this simple
model, but have yet to reach agreement regarding the strength of the external radiation 
field that is required. It is clear that the required radiation field strength is orders of
magnitude larger than the mean value produced by early generations of stars \citep{har00,gb06},
but it is also clear that the distribution of UV radiation in the early Universe is highly
inhomogeneous, owing to the strong clustering of the high-redshift protogalaxies
responsible for producing it \citep{dijk08,ahn09,agar12,dfm14}. The question of how rarely the 
required conditions are realized in the early Universe therefore depends sensitively on 
exactly how strong a UV field is actually required. This is usually quantified in terms of
the specific intensity of the radiation field at the Lyman limit. Following \citet{har00}, we
can measure this in units of $10^{-21} \: {\rm erg s^{-1} cm^{-2} Hz^{-1} sr^{-1}}$ and
write it as $J_{21}$. The minimum value of $J_{21}$ for which H$_2$ cooling is
sufficiently suppressed is then commonly denoted as $J_{\rm crit}$.

Values quoted in the literature for $J_{\rm crit}$ range from as low as 20 \citep{io11}
to as high as $10^{5}$ \citep{om01}. Typically, these values are determined by modelling
the cooling and collapse of gas within minihalos with $T_{\rm vir} > 10^{4} \: {\rm K}$ in
the presence of a strong UV background using 3D numerical simulations that treat the 
coupled thermal, chemical and dynamical evolution of the gas. By running the simulation
for a range of different values of $J_{21}$, it is possible to determine the critical value at which
H$_{2}$ cooling becomes sufficiently suppressed for the direct collapse model to operate.

The scatter in the values of $J_{\rm crit}$ obtained from these studies has several different
causes. $J_{\rm crit}$ depends strongly on the spectral shape adopted for the background radiation
field \citep{bl03,sbh10} and the manner in which H$_{2}$ self-shielding is accounted for \citep{wh11}, 
and also appears to vary somewhat from minihalo to minihalo \citep{sbh10,latif14}. 
However, an important additional contribution to the uncertainty in $J_{\rm crit}$ comes from the 
treatment of the gas chemistry within the different studies. 

In most cases, the chemical networks used in these studies were originally 
designed for the study of Pop.\ III star formation in the absence of a strong extragalactic radiation field. 
Although they all include the same basic processes (H$_{2}$ formation via the intermediate H$^{-}$ and
H$_{2}^{+}$ ions, H$_{2}$ destruction via collisional dissociation, charge transfer with H$^{+}$ and
photodissociation, and the photodissociation of H$^{-}$ and H$_{2}^{+}$), the exact set of chemical
reactions included varies from network to network. In addition, it is unclear whether {\em any} of the
networks in common usage includes the full set of chemical processes that are important for modelling 
the formation and  destruction of H$_{2}$ in an environment much harsher than the one that they 
were designed to model.

In this paper, we attempt to identify the key reactions that must be accounted for when modelling the
chemical evolution of the gas in this environment. To do this, we first model the chemical evolution of
the gas using an extensive model of primordial gas chemistry that tracks 30 different chemical species,
linked by almost 400 reactions. We then use a reaction-based reduction technique developed by 
\citet{Wiebe03} to analyze the chemistry and to identify the main reactions responsible for regulating 
the H$_{2}$ abundance. This allows us to construct a much smaller ``reduced'' network that contains
all of the chemical reactions that must be included in our chemical network if we are to be able to 
determine $J_{\rm crit}$ accurately.

The plan of the paper is as follows. In Section~\ref{num}, we present the details of the
reaction-based reduction technique and the one-zone chemistry and cooling model used 
to simulate the thermal and chemical evolution of the gas. In Section~\ref{results}, we 
present the results of our analysis and compare our reduced chemical network with 
other chemical networks used in the literature to simulate the formation of black holes
by direct collapse. Finally, we conclude in Section~\ref{sum} with a brief summary of our 
main results.

\section{Numerical method}
\label{num}
\subsection{Selecting the set of important  reactions}
\label{select}
In the direct collapse scenario, the crucial quantity that determines whether or not the gas is able
to cool significantly as it undergoes gravitational collapse is the abundance of molecular hydrogen.
If the amount of H$_{2}$ that forms does not provide sufficient cooling, then the gas remains warm
during the collapse, with the temperature $T$ decreasing only slowly as the density increases
\citep[see e.g.][]{om01,ssg10}. On the other hand, if enough H$_{2}$ forms to allow the gas to cool 
in less than a free-fall time, then the temperature drops dramatically once the density increases above 
$n \sim 10^{2}$--$10^{3} \: {\rm cm^{-3}}$. To determine the critical Lyman-Werner flux, $J_{\rm crit}$, one 
therefore simply needs to find the point at which the strength of the radiation field becomes
large enough to suppress the H$_{2}$ abundance sufficiently to prevent it from cooling the gas.

It follows from this that in order to determine $J_{\rm crit}$ accurately, we need to have an adequate
chemical model of the formation and destruction of H$_{2}$ in the gravitationally collapsing gas.
At the same time, however, we do not want to make our chemical model larger than is absolutely 
necessary, as each additional chemical species or chemical reaction that we include will further slow 
our numerical simulations. We therefore want to identify only those reactions and species that are most
important for modelling the evolution of the H$_{2}$ abundance in the collapsing gas.

To do this, we make use of the reaction-based reduction technique developed by \citet{Wiebe03} and
used by them to compute the abundance of carbon monoxide and the fractional ionization of the gas
in simulated molecular clouds. The basic idea underlying this technique is quite simple. Starting from
a prescribed set of initial conditions, we first compute the chemical and thermal evolution of the gas 
for a given value of $J_{21}$  using the one-zone model described in Section~\ref{one-zone} below. 
The chemical network used in this one-zone model is as extensive as we can reasonably make it, and
should therefore include all of the reactions that are likely to play a significant role in the H$_{2}$ 
chemistry. 

We next select our starting set of what \citet{Wiebe03} term ``important'' species, i.e.\
those whose abundances we are interested in following accurately. In the present case, this set
consists of only a single member -- the H$_{2}$ molecule -- but in principle the same technique can be 
used with a larger set of important species. We then proceed by determining the dominant production
and destruction reactions for each of these species at a set of different output times during the
evolution of the gas. If the set of dominant reactions involve chemical species
that are not in our starting set, then we add them to the set of important species,
and repeat the analysis, proceeding in this fashion until there are no more 
species or reactions that need to be added. The result is a list of the reactions
that are necessary at each output time in order to accurately model the
abundances of our starting set of species. The make-up of this list will generally
change with time: some reactions that are important at early times will be 
unimportant at late times, and vice versa. The final set of important reactions 
can then be obtained simply by combining the individual lists.

To determine the dominant production and destruction reactions at any given
output time, we proceed iteratively, as follows. On the first iteration, we set the 
weights $w_{s}$ of the important species to 1, and the weights of all other
species to 0. We also set the weight of each reaction to zero. We then loop
over all of the chemical species included in our chemical model. For each
species $k$, we compute a new weight for each reaction involving that species:
\begin{equation}
w_{r}^{\rm new}(j) =  {\rm max}\left \{ w_{r}^{\rm curr}(j), \frac{|R_{j}|}{\sum_{l = 1, N_{r}(k)} R_{l}}
w^{\rm curr}_{s}(k) \right \}. \label{weight_eq}
\end{equation}
Here, $w^{\rm curr}_{s}(k)$ is the current weight of species $k$,  $w^{\rm curr}_{r}(j)$ is the
current weight of reaction $j$, $w^{\rm new}_{r}(j)$ is the new weight of reaction $j$, 
$R_{j}$ is the rate per unit volume at which reaction $j$ proceeds, and $N_{r}(k)$ is the
number of reactions in which species $k$ participates as either a reactant or a product.
Once we have calculated a new weight for every reaction involving species $k$, we update
its weight:
\begin{equation}
w_{s}^{\rm new}(k) = {\rm max} \left[w_{s}^{\rm curr}(k), {\rm max} \left \{ w_{r}^{\rm new}(1), ..., 
w_{r}^{\rm new}(N_{r}(k)) \right \} \right].
\end{equation}
In other words, we take the new species weight to be the largest value from amongst the
weights of the reactions involving that species, unless this is smaller than the current weight
of the species. Having determined new weights for every species, we begin a new iteration,
using the updated weights in Equation~\ref{weight_eq} above. We continue to iterate until
all of the species weights have converged to within some pre-specified tolerance. 

One situation in which this procedure can give misleading results occurs when the formation
and destruction of one of the species is dominated by a single forwards reaction and its 
inverse. Consider, for example, the case of the formation of HD from H$_{2}$
\begin{equation}
{\rm H_{2} + D^{+}} \rightarrow {\rm HD + H^{+}}.
\end{equation}
In warm gas, this reaction occurs very rapidly, but so does its inverse
\begin{equation}
{\rm HD + H^{+}} \rightarrow {\rm H_{2} + D^{+}}.
\end{equation}
In the extreme case in which these reactions are perfectly balanced, both can have large
$R$ values, and hence potentially also large weights, and yet the net effect of the pair of
reactions is to leave the abundances of all of the species involved unchanged. In our analysis,
we attempt to mitigate the impact of this problem by identifying reaction pairs of this kind and
replacing the reaction rates used for these reactions in Equation~\ref{weight_eq} with the
absolute value of the difference between the forward and reverse reaction rates; in the example
above, this corresponds to the net rate at which H$_{2}$ is formed or destroyed by the pair
of reactions.

Our analysis procedure leaves us with a list of weights for every reaction that is valid for the 
output time considered. We can convert this into a list of reactions by retaining only those 
reactions whose weights exceed some cut-off value $\epsilon$. As we show later
in Section~\ref{epsilon}, setting $\epsilon = 10^{-4}$ enables us to generate a subset of reactions that
allow us to determine  $J_{\rm crit}$ to within an accuracy of around 1\% when compared 
to the results of models run with the full chemical network. 

By repeating the same
procedure for many different output times, and combining the individual reaction lists, we
can construct a reduced reaction network that is sufficient for modelling the chemical 
evolution of H$_{2}$ over the entire period simulated in our one-zone model. In principle,
the reduced reaction network that we derive in this fashion is specific to our choice of
$J_{21}$ and to the initial conditions for our simulation. However, by re-running the 
one-zone model and repeating the same analysis procedure for many different values 
of $J_{21}$ with various different sets of initial conditions, as described in Section~\ref{one-zone}
below, and the combining the resulting reduced networks, we can arrive at a
final reduced network that is valid over the whole range of physical conditions that are
of interest to us.

\subsection{The one-zone model}
\label{one-zone}
\subsubsection{Basic details}
In order to model the chemical and thermal evolution of gravitationally-collapsing primordial
gas illuminated by a strong extragalactic radiation field, we use a simple one-zone model
that is derived from the one presented in \citet{Glover09}. The gas density is assumed to
evolve as
\begin{equation}
\frac{{\rm d}\rho}{{\rm d}t} = \frac{\rho}{t_{\rm ff}},
\end{equation}
where $t_{\rm ff} = (3\pi / 32 G \rho)^{1/2}$ is the free-fall time of the gas. To follow the
thermal evolution of the gas, we write the internal energy equation as
\begin{equation}
\frac{{\rm d}e}{{\rm d}t} = \frac{p}{\rho^{2}} \frac{{\rm d}\rho}{{\rm d}t} + \Gamma - \Lambda,
\end{equation}
where $e$ is the internal energy density, $p = (\gamma - 1)e$ is the gas pressure, 
$\Gamma$ is the radiative and chemical heating rate per unit volume and $\Lambda$ is the 
radiative and chemical cooling rate per unit volume. 

We compute $\Gamma$ and $\Lambda$ using a detailed atomic and molecular cooling
function based on the one presented in Section~4 of \citet{Glover09}. In an effort to use
the most up-to-date atomic and molecular data available, we have updated two of the
cooling processes included in the model: the collisional excitation of H$_{2}$ by collisions
with protons and by collisions with electrons. Full details of these updates are given in Appendix~\ref{app:cool}.

To model the chemical evolution of the gas, we use a modified version of the chemical
network present in \citet{Glover09}. The original version of this network tracked the
abundance of 30 different primordial chemical species and included a total of 392
reactions. We have extended this model by including three new reactions not treated 
in \citet{Glover09}: the collisional ionization of atomic hydrogen by collisions with H or He,
and the collisional dissociation of H$_{2}^{+}$ by H. Details of these new reactions are
given in Appendix~\ref{app:chem}. We have also updated several of the reaction rates
in light of new theoretical or experimental data. Again, full details of these modifications
are given in Appendix~\ref{app:chem}.

\subsubsection{Photochemistry and self-shielding}
The rate coefficients listed in \citet{Glover09} for the various photochemical reactions assumed
that the external background radiation field had the spectral shape of a $10^{5}$~K black-body,
cut-off at energies of 13.6~eV and above to account for the effects of absorption by atomic
hydrogen in the intergalactic medium. This choice of spectrum (which we refer to hereafter
as a T5 spectrum, for brevity) was motivated by the fact that we expect the brightest Pop.\ III stars
to have effective temperatures of around this value \citep[see e.g.][]{coj00}. However, this is a
reasonable approximation only when the dominant contribution to the extragalactic background
radiation field does indeed come from very massive Pop.\ III stars, specifically those with masses
greater than around $100 \: {\rm M_{\odot}}$ \citep{bkl01}. If a significant fraction of the radiation
comes instead from less massive Pop.\ III stars or from Pop.\ II stars with systematically lower
effective temperatures, then the ratio of the UV to the optical flux can potentially be much smaller
than with a T5 spectrum, with important implications for the relative importance of H$_{2}$
photodissociation and H$^{-}$ photodetachment. Several authors have attempted to quantify
the dependence of $J_{\rm crit}$ on the spectral shape of the extragalactic radiation field
by performing simulations both with a T5 spectrum and with a T4 spectrum: a $10^{4}$~K 
diluted black-body with a similar high-energy cut-off to the T5 spectrum. These studies
\citep[see e.g.][]{bl03,sbh10} find that the choice of spectrum has a large influence on the value 
of  $J_{\rm crit}$. For instance, \citet{sbh10} find that in their models, the value of $J_{\rm crit}$
for a T5 spectrum lies in the range $10^{4} < J_{\rm crit} < 10^{5}$, but that for a T4 spectrum,
$30 < J_{\rm crit} < 300$, a difference of two orders of magnitude or more.

To account for this uncertainty, we perform two sets of simulations: one set using 
a T5 spectrum and a second set using a T4 spectrum. For the runs with the T5
spectrum, we use the photochemical rates listed in \citet{Glover09} for all processes
except for the photodissociation of H$_{2}^{+}$, which is treated with the density-dependent 
approach described in Appendix~\ref{updated}. For the other set of runs, we have
recomputed the photochemical rates using a T4 spectrum and the same basic
approach as in \citet{Glover09}.

In practice, of course, both the T4 and T5 spectra are relatively crude approximations
to the actual spectrum of the extragalactic background produced by high-redshift 
star-forming protogalaxies \citep{ak14,soi14,latif15}. However, they should bracket
the behaviour seen in more realistic models, making them suitable choices for the
purposes of our study.

When treating most of the photochemical reactions, we assume that the gas remains
optically thin, as the continuum opacity of low density primordial gas is very small
\citep[see e.g.][]{lcs91}. The important exception is H$_{2}$ photodissociation, as the rate of this processes
can be strongly affected by H$_{2}$ self-shielding. To account for this, 
we use the modified form of the
\citet{db96} shielding function introduced by \citet{whb11}:
\begin{eqnarray}
f_{\rm sh}(N_{\rm H_{2}}, T) & = & \frac{0.965}{(1 + x/b_{5})^{1.1}} + \frac{0.035}{(1+x)^{0.5}} \nonumber \\
& & \times \exp \left[-8.5 \times 10^{-4} (1+x)^{0.5} \right],
\end{eqnarray}
where $x = N_{\rm H_{2}} / 5 \times 10^{14} \: {\rm cm^{-2}}$, $b_{5} = b / 10^{5} \: {\rm cm \: s^{-1}}$,
and $b$ is the Doppler broadening parameter, which we assume to be dominated by the effects
of thermal broadening.
In order to evaluate this expression for $f_{\rm sh}$, we need to know the column density of H$_{2}$ in the
collapsing gas cloud. We compute this using a similar method to \citet{om01}: we assume that 
the dominant contribution to the shielding comes from a core region with radius 
$R_{\rm c} = \lambda_{\rm J}$, where $\lambda_{\rm J}$ is the Jeans length.
The required H$_{2}$ column density then follows as $N_{\rm H_{2}} = n_{\rm H_{2}} R_{\rm c}
= n_{\rm H_{2}} \lambda_{\rm J}$, where $n_{\rm H_{2}}$ is the number density of
H$_{2}$ in our one-zone model. 

It should be noted that the simple approximation that we use here to compute $N_{\rm H_{2}}$
has been shown to overestimate the actual amount of shielding by up to an
order of magnitude in some cases compared to the results of fully 3D treatments \citep{whb11}.
For this reason, the absolute values of $J_{\rm crit}$ that we derive in this study should be treated
with considerable caution. However, this simplification should not affect our conclusions regarding
the set of chemical reactions that are required to accurately determine $J_{\rm crit}$, or our
findings regarding the sensitivity of $J_{\rm crit}$ to uncertainties in the rates of these reactions.

\subsubsection{Initial conditions}
We perform simulations using three different set of initial conditions, as detailed in Table~\ref{tab:sims}.
In runs 1 and 4, we start with a low initial temperature, $T_{0} = 200$~K, and an initial fractional
ionization and H$_{2}$ fractional abundance close to those found in the IGM prior to the onset of
Pop.\ III star formation ($x_{\rm e, 0} = 2 \times 10^{-4}$, $x_{\rm H_{2},0} = 2 \times 10^{-6}$).
In runs 2 and 5, we start with the same chemical composition but with a much higher gas
temperature, $T_{0} = 8000$~K. Finally, in runs 3 and 6, we set $T_{0} = 8000$~K, but assume
that the gas was initially fully ionized ($x_{\rm e, 0} = 1.0$, $x_{\rm H_{2}, 0} = 0.0$). The initial
density in all of the simulations is set to $n_{0} = 0.3 \: {\rm cm^{-3}}$, corresponding approximately
to the mean density within the virial radius of a protogalaxy forming at a redshift $z = 20$. We have explored
the effects of adopting a larger initial density but find that this makes little difference to our results.

In all of our simulations, we adopt an elemental abundance (relative to hydrogen) of 
$A_{\rm D} = 2.6 \times 10^{-5}$ for deuterium and $A_{\rm Li} = 4.3 \times 10^{-10}$ for 
lithium \citep{cyb04}. We set the initial abundances of D$^{+}$ and HD so that they are
a factor of $A_{\rm D}$ smaller than the initial H$^{+}$ and H$_{2}$ abundances, respectively.
We assume that the lithium was initially entirely in neutral atomic form, and set the initial
abundances of all of the other chemical species to zero.
 
For each set of initial conditions, we run two different sets of simulations, one with a T4 spectrum
and the other with a T5 spectrum, as indicated in Table~\ref{tab:sims}.
  
\begin{table}
\caption{List of simulations \label{tab:sims}}
\begin{tabular}{ccccc}
\hline
Run & $T_{0}$ (K) & $x_{\rm e, 0}$ & $x_{\rm H_{2}, 0}$ & Spectrum  \\
\hline
1 & 200   & $2 \times 10^{-4}$ & $2 \times 10^{-6}$ & T4  \\
2 & 8000 & $2 \times 10^{-4}$ &$2 \times 10^{-6}$ & T4  \\
3 & 8000 & $1.0$ & 0.0 & T4 \\
4 & 200   & $2 \times 10^{-4}$ & $2 \times 10^{-6}$ & T5 \\
5 & 8000 & $2 \times 10^{-4}$ & $2 \times 10^{-6}$ & T5 \\
6 & 8000 & $1.0$ & 0.0 & T5 \\
\hline
\end{tabular}
\end{table}

\section{Construction of an accurate reduced network}
\label{results}
\subsection{Determination of $J_{\rm crit}$}
Before we can identify the set of reactions that it is necessary to include in the chemical model in order to accurately
determine $J_{\rm crit}$, it is first necessary to establish the value of $J_{\rm crit}$ for each of our six different setups.
We do this using a binary search method. We start by selecting two values of $J_{21}$ that are certain to bound
$J_{\rm crit}$: a low value, $J_{\rm 21, low} = 1$, that we have verified is too small to prevent efficient H$_{2}$ cooling from
occurring in any of the models, and a high value, $J_{\rm 21, high} = 10^{4}$, that completely suppresses H$_{2}$ cooling 
in all of the models. We then compute a new value of $J_{21}$ using the equation
\begin{equation}
J_{\rm 21, new} = \left(J_{\rm 21, low} \times  J_{\rm 21, high} \right)^{1/2},
\end{equation}
and run a simulation using this new value. If $J_{\rm 21, new}$ is large enough to prevent efficient H$_{2}$ cooling, then 
we adopt it as our new value of $J_{\rm 21, high}$; otherwise, we take it as our new value of $J_{\rm 21, low}$. We 
proceed in this fashion until the difference between $J_{\rm 21, low}$ and $J_{\rm 21, high}$ is less than 0.2\%,
and take the final value of $J_{\rm 21, new}$ as our estimate for $J_{\rm crit}$. The resulting values are listed in Table~\ref{tab:jcrit}
for each of our six different setups. In agreement with previous work, we find that the choice of spectral shape has a large
influence on $J_{\rm crit}$. With a T4 spectrum, we find that $J_{\rm crit} \sim 17$--18, while a T5 spectrum yields $J_{\rm crit} \sim 1630$.
We see also that the results that we obtain are not particularly sensitive to our choice of initial conditions.


\begin{table}
\caption{Critical Lyman-Werner flux as a function of $\epsilon$ \label{tab:jcrit}}
\begin{tabular}{cccccc}
\hline 
&  \multicolumn{5}{c}{$J_{\rm crit}$} \\
Run & $\epsilon = 0$ & $\epsilon = 10^{-4}$ & $\epsilon = 10^{-3}$ & $\epsilon = 0.01$ & $\epsilon = 0.1$ \\
\hline
1 & 17.0 & 17.1 & 17.1 & 17.1 & 17.5 \\
2 & 18.0 & 18.0 & 18.0 & 18.1 & 18.6 \\
3 & 18.0 & 18.1 & 18.1 & 18.1 & 18.6 \\
4 & 1640 & 1640 & 1640 & 1650 & 4260 \\
5 & 1630 & 1630 & 1630 & 1640 & 4260 \\  
6 & 1630 & 1630 & 1630 & 1640 & 4260 \\
\hline
\end{tabular}
The values shown here are specified to only three significant figures, and
were computed using chemical networks consisting of all chemical reactions with
maximum weights exceeding the listed value of $\epsilon$. The case $\epsilon = 0$ corresponds
to the full chemical model.
\end{table}

It is interesting to compare our results to those of previous one-zone studies. For the T4 spectrum, our finding that $J_{\rm crit} \sim 17$--18 is in fairly good agreement with the values $J_{\rm crit} = 20, 25, 39$ derived by \citet{io11}, \citet{soi14}, and \citet{sbh10}, respectively. The difference between our value and the values derived in these other studies is most likely due to the impact of the differences in the chemical model and the choices made for some of the key rate coefficients, since as we will see later in this paper and in Paper II 
\citep{glover15}, these differences can easily introduce an uncertainty of a factor of a few into our estimates for $J_{\rm crit}$.

For the T5 spectrum, our value of $J_{\rm crit}$ is almost an order of magnitude smaller than the values of $J_{\rm crit} = 1.2 \times 10^{4}$ and $1.6 \times 10^{4}$ found by \citet{sbh10} and \citet{io11}, respectively. However, this difference can be explained by differences in our treatment of H$_{2}$ self-shielding. Both of these previous studies used the H$_{2}$ self-shielding function computed by \citet{db96}, while we used instead the modified version given in \citet{whb11}, which more accurately treats the self-shielding of H$_{2}$ in hot gas. \citet{soi14} performed one-zone calculations with a T5 spectrum using both of these treatments of H$_2$ self-shielding, and showed that the value of $J_{\rm crit}$ that they obtained with the  \citet{whb11} treatment was approximately an order of magnitude smaller than that obtained using the \citet{db96} treatment. The value they obtained for $J_{\rm crit}$ with the  \citet{whb11} treatment was $J_{\rm crit} \simeq 1400$, in good agreement with the value we find using our model.

\subsection{The reduced network}
\label{reduce}
Having determined the value of $J_{\rm crit}$ in each run, we next analyze the full set of chemical reactions taking place during the evolution of the gas, as outlined in Section~\ref{select}. For each run, we consider a large set of output times, and use our computed values for the density, temperature and chemical composition of the gas to determine the weight of each reaction in our full set at that output time. We record these weights, and repeat this procedure for a large number of different output times. We then determine the maximum weight for each reaction in our model. To help ensure that our reduced network will be robust against minor changes in the physical conditions in the gas, we consider not only the case when $J_{21} = J_{\rm crit}$, but also perform the same analysis for simulations with $J_{21} = 0.3 J_{\rm crit}$ and $J_{21} = 3 J_{\rm crit}$, taking the maximum weight for each reaction to be the largest of the weights obtained for the reaction with these three different setups. Finally, we construct our reduced network using only those reactions whose maximum weights exceed $\epsilon = 10^{-4}$. The resulting set of reactions for each run is listed in Table~\ref{tab:reduce}, along with the corresponding set of maximum weights.

\begin{table*} 
\caption{List of reactions with maximum reaction weights greater than $10^{-4}$ in at least one simulation \label{tab:reduce}}
\begin{tabular}{clcccccc}
\hline
& & \multicolumn{6}{c}{Maximum reaction weight} \\
No.\ & Reaction &  Run  1 & Run 2 & Run 3 & Run 4 & Run 5 & Run 6 \\
\hline
{\bf 1} & ${\rm H_{2} + \gamma} \rightarrow {\rm H + H}$ & 1.00 & 1.00 & 1.00 & 1.00 & 1.00 & 1.00 \\
{\bf 2} & ${\rm H_2 + H} \rightarrow {\rm H + H + H}$ & 0.49 & 0.50 & 0.49 & 0.49 & 0.49 & 0.49 \\
{\bf 3} & ${\rm H^{-} + H} \rightarrow  {\rm H_{2} + e^{-}}$ & 0.47 & 0.47 & 0.47 & 0.50 & 0.47 & 0.47 \\
{\bf 4} & ${\rm H_{2}^{+} + H} \rightarrow {\rm H_{2} + H^{+}}$ & 0.41 & 0.49 & 0.48 & $4.7 \times 10^{-2}$ &$4.3 \times 10^{-2}$ & 
$4.4 \times 10^{-2}$ \\
{\bf 5} & ${\rm H^{+} + e^{-}} \rightarrow {\rm H + \gamma}$ & 0.27 & 0.12 & 0.48 & $4.7 \times 10^{-3}$ & $1.0 \times 10^{-2}$ & $4.3 \times 10^{-2}$ \\
{\bf 6} & ${\rm H + e^{-}} \rightarrow {\rm H^{-} + \gamma}$ & 0.23 & 0.23 & 0.23 & 0.25 & 0.23 & 0.24 \\
{\bf 7} & ${\rm H^{-} + \gamma} \rightarrow {\rm H + e^{-}}$ & 0.21 & 0.15 & 0.14 & 0.25 & 0.23 & 0.23 \\
{\bf 8} & ${\rm H + H^{+}} \rightarrow {\rm H_{2}^{+} + \gamma}$ & 0.20 & 0.24 & 0.24 & $2.3 \times 10^{-2}$ & $2.2 \times 10^{-2}$  
& $2.1 \times 10^{-2}$ \\
{\bf 9} & ${\rm H_{2}^{+} + \gamma} \rightarrow {\rm H^{+} + H}$ & 0.12 & 0.24 & 0.24 & $7.2 \times 10^{-3}$ & $2.1 \times 10^{-2}$ & 
$2.0 \times 10^{-2}$ \\
{\bf 10} & ${\rm H + H} \rightarrow {\rm H^{+} + e^{-} + H}$ & $9.5 \times 10^{-2}$ & 0.12 & $1.1 \times 10^{-2}$ & $1.4 \times 10^{-2}$ & $1.1 \times 10^{-2}$ & $1.4 \times 10^{-3}$ \\
{\bf 11} & ${\rm H^{-} + H} \rightarrow {\rm H + H + e^-}$ & $8.8 \times 10^{-2}$ & $8.9 \times 10^{-2}$ &  $8.8 \times 10^{-2}$ & $9.3 \times 10^{-2}$ & $9.2 \times 10^{-2}$ & $9.2 \times 10^{-2}$ \\
{\bf 12} & ${\rm H + e^{-}}  \rightarrow {\rm H^+ + e^- + e^-}$ & $6.1 \times 10^{-2}$ & 0.22 & $4.0 \times 10^{-2}$ & $7.8 \times 10^{-3}$ & $2.0 \times 10^{-2}$ & $3.9 \times 10^{-3}$ \\
{\bf 13} & ${\rm H_{2}^{+} + He}  \rightarrow {\rm HeH^+ + H} $ & $3.5 \times 10^{-2}$  & $2.8 \times 10^{-3}$ & $2.7 \times 10^{-3}$  & $3.6 \times 10^{-4}$ & $3.0 \times 10^{-4}$ & $3.0 \times 10^{-4}$ \\
{\bf 14} & ${\rm H + He}  \rightarrow {\rm H^+ + e^- + He} $ & $2.7 \times 10^{-2}$ & $3.6 \times 10^{-2}$ & $3.1 \times 10^{-3}$ & $3.3 \times 10^{-3}$ & $3.2 \times 10^{-3}$ & $3.9 \times 10^{-4}$ \\
{\bf 15} & ${\rm H_2 + H^+}  \rightarrow {\rm H_2^+ + H} $ &   $1.0 \times 10^{-2}$ & $8.0 \times 10^{-3}$ & $4.2 \times 10^{-2}$ & $1.1 \times 10^{-3}$ & $1.1 \times 10^{-3}$ & $1.1 \times 10^{-3}$ \\
16 & ${\rm H_2 + He}  \rightarrow {\rm H + H + He} $ & $5.9 \times 10^{-3}$ & $5.9 \times 10^{-3}$ & $5.9 \times 10^{-3}$ & $5.8 \times 10^{-3}$ & $5.8 \times 10^{-3}$ & $5.8 \times 10^{-3}$ \\
{\bf 17} & ${\rm HeH^{+} + H}  \rightarrow {\rm H_2^+ + He}$ &  $2.3 \times 10^{-3}$& $2.8 \times 10^{-3}$ & $2.6 \times 10^{-3}$ & $3.6 \times 10^{-4}$ & $3.0 \times 10^{-4}$ & $3.0 \times 10^{-4}$ \\
18 & ${\rm H + H + H}  \rightarrow {\rm H_2 + H } $ & $1.4 \times 10^{-3}$ & $1.3 \times 10^{-3}$ & $1.3 \times 10^{-3}$ & --- & --- & --- \\
19 & ${\rm H^{-} + He}  \rightarrow {\rm H + He + e^-} $ & $1.3 \times 10^{-3}$ & $1.4 \times 10^{-3}$ & $1.3 \times 10^{-3}$ & $2.5 \times 10^{-3}$ & $2.0 \times 10^{-3}$ & $1.9 \times 10^{-3}$ \\
20 & ${\rm H_{2}^{+} + H} \rightarrow {\rm H + H^{+} + H}$ & $1.3 \times 10^{-3}$ & $3.6 \times 10^{-4}$ &$5.2 \times 10^{-4}$ & --- & --- & ---  \\
21 & ${\rm He + H^{+}}  \rightarrow {\rm HeH^{+} + \gamma}$ & $5.3 \times 10^{-4}$ & --- & --- & --- & --- & --- \\
{\bf 22} & ${\rm H^{-} + {\rm H^{+}}} \rightarrow {\rm H + H}$ & $4.6 \times 10^{-4}$  & $4.6 \times 10^{-4}$ & $4.0 \times 10^{-4}$ & $7.5 \times 10^{-4}$ & $1.4 \times 10^{-3}$ & $6.6 \times 10^{-2}$ \\
{\bf 23} & ${\rm H_{2}^{+} + e^{-}} \rightarrow {\rm H + H}$ & $1.7 \times 10^{-4}$ & $2.0 \times 10^{-4}$ & $2.4 \times 10^{-2}$ & --- & --- & $8.7 \times 10^{-4}$ \\
24 & ${\rm HeH^{+} + e^{-}} \rightarrow {\rm He + H}$ & --- & --- & $1.9 \times 10^{-4}$ & --- & --- & --- \\
25 & ${\rm H^{-} + H^{+}} \rightarrow {\rm H_{2}^{+} + e^{-}}$ & --- & --- & $1.0 \times 10^{-4}$ & --- & --- & $9.6 \times 10^{-4}$ \\
26 & ${\rm H^{-} + e^{-}} \rightarrow {\rm H + e^{-} + e^{-}}$ & --- & --- & --- & $1.0 \times 10^{-4}$ & $2.6 \times 10^{-4}$ & $9.5 \times 10^{-3}$ \\
\hline
\end{tabular}
\\ The reactions with numbers listed in bold font make up the minimal reduced model described in Section~\ref{epsilon}.
\end{table*}

We see immediately that there is a subset of reactions with very large weights that appear in the reduced network for each of our runs. This is not surprising: most of the reactions in this set play such important roles in the regulation of the H$_{2}$ abundance that any model omitting them could not hope to be representative of the real behaviour of the gas. Examples include the photodissociation and collisional dissociation of H$_{2}$, the formation of H$_{2}$ from H$^{-}$, or the formation and photodissociation of the H$^{-}$ ion itself. 

There are also a few reactions that have relatively high weights in some of our runs that are more unexpected. Perhaps the best example of these is the collisional ionization of atomic hydrogen by collisions with other hydrogen atoms, i.e.\
\begin{equation}
{\rm H + H} \rightarrow {\rm H + H^{+} + e^{-}}.  
\end{equation}
Most previous networks developed to treat primordial chemistry have omitted this reaction\footnote{The notable exception is \citet{om01}, who does include this process in his chemical network. Studies of SMBH formation that make use of his network \citep[e.g.][]{soi14} therefore already account for this reaction.}, assuming it to be unimportant in comparison to the collisional ionization of hydrogen by electrons, since the latter reaction has a far larger rate coefficient. However, while it is certainly true that collisions with electrons dominate in many circumstances -- for example in gas cooling and recombining from an initially ionized state -- this is not true in all of the runs we examine here. In particular, if the initial electron abundance is small, as for example in runs 1 or 2, then collisions between hydrogen atoms happen so much more frequently than collisions between hydrogen atoms and electrons that this reaction can come to dominate the ionization rate despite its small rate coefficient. 

To illustrate the importance of this reaction, we re-ran our determination of $J_{\rm crit}$ for models 1 and 4 with the value of its rate coefficient set to zero. We found that in this case $J_{\rm crit} = 11.8$ for run 1 and $J_{\rm crit} = 1280$ for run 4. In other words, omitting this reaction leads to a 20--30\% decrease in the derived value of $J_{\rm crit}$ when we start from low ionization initial conditions.  If we also disable the reaction
\begin{equation}
{\rm H + He} \rightarrow {\rm H^{+} + e^{-} + He},
\end{equation}
which again is often omitted from treatments of primordial chemistry, then $J_{\rm crit}$ is reduced even further, to $J_{\rm crit} = 8.9$ or 1110 for runs 1 and 4, respectively.

\begin{figure}
\includegraphics[width=3.2in]{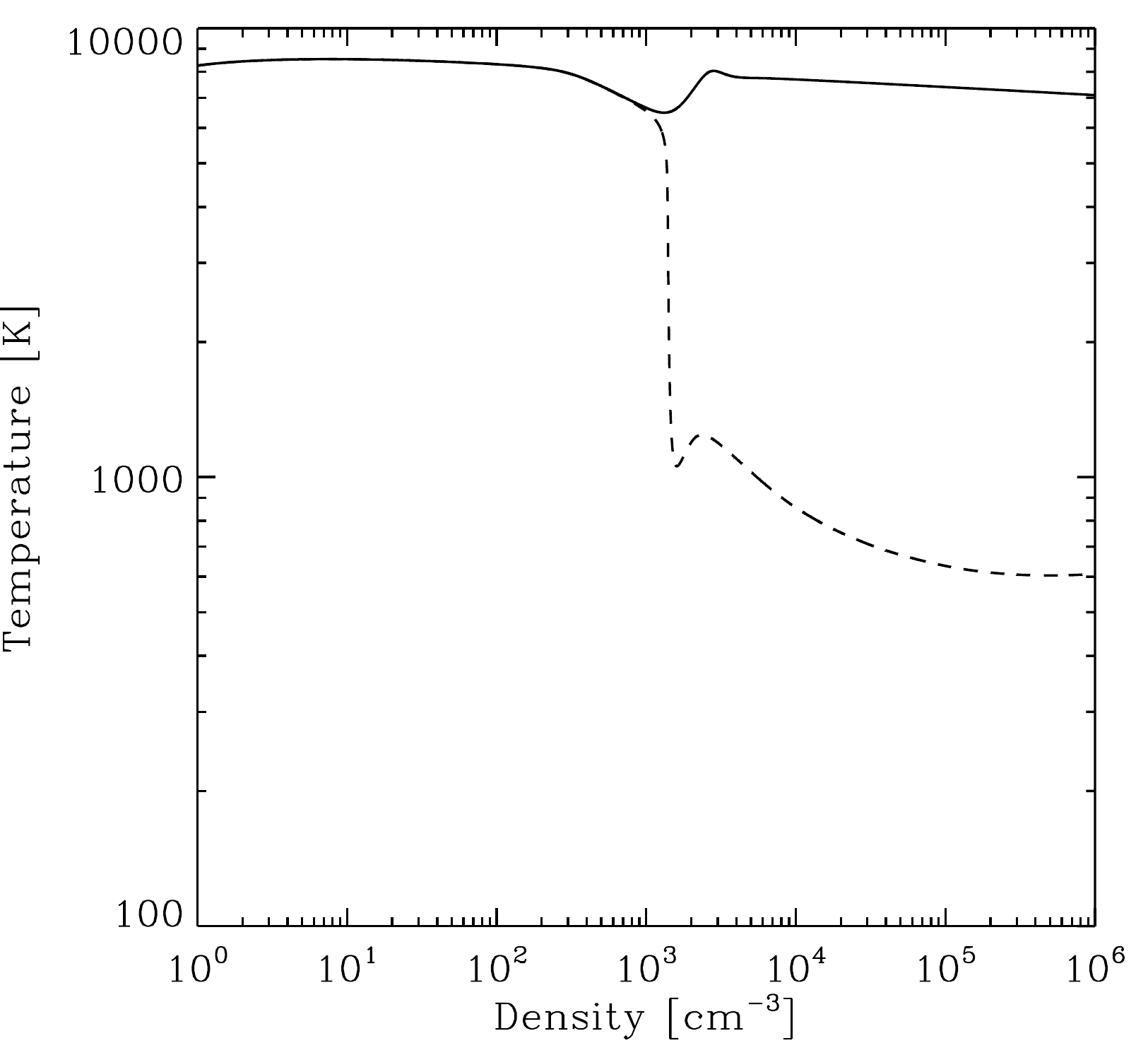}
\includegraphics[width=3.2in]{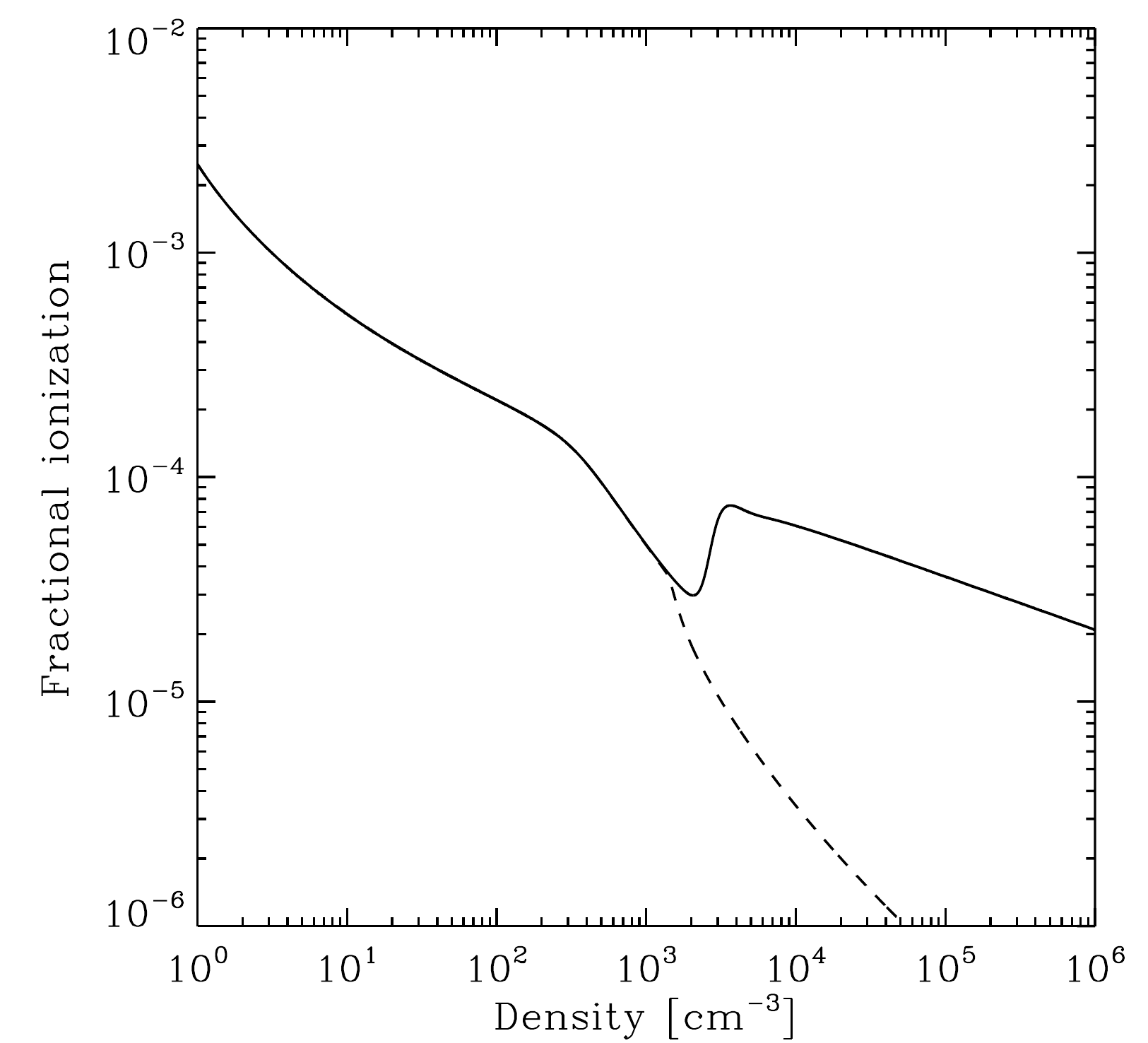}
\caption{{\em Top panel}: evolution of the gas temperature as a function of the hydrogen nuclei number density in runs with the same setup as in run 6 (a T5 spectrum, fully ionized initial conditions). Results are shown for $J_{21} = 1630$ (solid line) and $J_{21} = 1635$ (dashed line). 
{\em Bottom panel}: evolution of the fractional ionization as a function of density in the same runs.
\label{evol}}
\end{figure}

We can understand why these two reactions appear to be so important for determining $J_{\rm crit}$ if we examine how the fractional ionization of the gas evolves as a function of density when $J_{21}$ is close to $J_{\rm crit}$. In Figure~\ref{evol}, we show how the temperature and the fractional ionization evolve with density in two simulations carried out with the same initial conditions as run 6 (i.e.\ fully ionized gas, illuminated by a T5 spectrum), with $J_{21} = 1630$ (dashed line) and $J_{21} = 1635$ (solid line). These two values of $J_{21}$ were chosen to tightly bracket $J_{\rm crit}$. We see that in both simulations, the temperature evolution is essentially identical until we reach a density of $n \sim 10^{3} \: {\rm cm^{-3}}$. At this point, the behaviour of the two runs diverges. In the run with $J_{21} < J_{\rm crit}$, the H$_2$ fraction in the gas at this point is high enough to allow it to begin to cool significantly. As it cools, the effects of H$_{2}$ collisional dissociation become less effective, allowing the H$_{2}$ fraction to increase further, and so the cooling runs away, rapidly driving the temperature down to $T \sim 1000$~K. In the run with $J_{21} > J_{\rm crit}$, on the other hand, the smaller H$_{2}$ abundance means that H$_{2}$ cooling never becomes effective and the gas remains hot throughout the simulation.

The value of $J_{\rm crit}$ is therefore determined in this case by the chemical state of the gas at $n \sim 10^{3} \: {\rm cm^{-3}}$ and $T \sim 7500$~K. The H$_{2}$ fraction in the gas in these conditions depends on the fractional ionization. Figure~\ref{evol} shows that even though the gas is initially fully ionized, by the time we reach $n \sim 10^{3} \: {\rm cm^{-3}}$, the fractional ionization has lost its memory of this initial state and has decreased to $x \sim 4 \times 10^{-5}$. This value is set by the balance between the collisional ionization of hydrogen by collisions with electrons, H atoms and He atoms (i.e.\ reactions 12, 10, and 14, respectively) and the radiative recombination of H$^{+}$. If we compare the rates of reactions 10, 12 and 14 in these conditions, we find that $R_{10} \simeq 4 \times 10^{-16} \: {\rm cm^{-3} \: s^{-1}}$, $R_{12} \simeq 2 \times 10^{-16} \: {\rm cm^{-3} \: s^{-1}}$ and $R_{14} \simeq 10^{-16} \: {\rm cm^{-3} \: s^{-1}}$. In other words, because of the very low fractional ionization of the gas, reactions 10 and 14 between them provide around 70\% of the total collisional ionization rate, with the main contribution coming from reaction 10. Omitting these reactions therefore leads to a factor of $\sim 3$ error in the total ionization rate, and hence a factor of $\sim 2$ error in the fractional ionization. This reduces the H$_{2}$ formation rate by a similar amount and hence also reduces our estimate of $J_{\rm crit}$.

Another interesting thing that Table~\ref{tab:reduce} shows us is that once we start to look at reactions with lower weights, which are less important overall for our determination of $J_{\rm crit}$, we see substantially more variation from run to run. Only for runs 4 and 5 does our reaction weighting algorithm provide us with exactly the same set of reactions with maximum weights greater than $\epsilon$; for the other runs, we see minor differences, generally involving a small number of processes with weights close to our cut-off. In order to have a reaction network that is robust to changes in the initial conditions and details of the background radiation field, we therefore recommend that one uses all of the 26 reactions listed in Table~\ref{tab:reduce}.

\subsection{Effect of varying $\epsilon$}
\label{epsilon}
In order to establish that our choice of a value of $\epsilon = 10^{-4}$ for the reaction weight cut-off in our selection algorithm does indeed allow us to select all of the reactions important for determining $J_{\rm crit}$, we re-ran our models using a reduced chemical network consisting only of the 26 reactions listed in Table~\ref{tab:reduce} and recalculated $J_{\rm crit}$ for each model using the same approach as before. The resulting values are shown in Table~\ref{tab:jcrit}. We see that there are minor differences in the predicted values of $J_{\rm crit}$ in a couple of the runs, but that they are at the level of less than 1\%. This is much smaller than the scatter in $J_{\rm crit}$ from halo to halo caused by differences in the dynamical evolution of the gas \citep{latif14}.
This demonstrates that our 26 reaction reduced network is indeed accurate enough for our purposes. 

The question then naturally arises as to whether we can simplify our reduced network even more. If we increase the value of $\epsilon$, and hence select even fewer reactions to join the set of ``important'' reactions, then how badly does this compromise our ability to determine $J_{\rm crit}$ accurately? To explore this, we constructed several additional reduced networks, in which we retained only those reactions with maximum weights  greater than $\epsilon = 10^{-3}$, 0.01 or 0.1 in at least one run, and then used these even more highly simplified networks to determine $J_{\rm crit}$. When performing this analysis for the $\epsilon = 0.01$ case, we found that in order to produce meaningful results, it was necessary to include one reaction -- the destruction of HeH$^{+}$ by collisions with H, reaction 17 in the Table -- that formally has a weight that is less than 0.01. The reason for this is that otherwise our $\epsilon = 0.01$ model would contain a reaction that forms HeH$^{+}$ (reaction 13 in the Table), but no reaction that destroys it, meaning that the abundance of HeH$^{+}$ would increase indefinitely. 

The results that we obtained when varying $\epsilon$ are shown in columns 4-6 of Table~\ref{tab:jcrit}. We see that the results we obtain for $\epsilon = 10^{-3}$ or $\epsilon = 0.01$ are very similar to those we obtain in the $\epsilon = 10^{-4}$ case, with the derived values of $J_{\rm crit}$ differing by no more than around 1\%. However,  in the $\epsilon = 0.1$ case, we see a clear change in behaviour: the error in the value of $J_{\rm crit}$ we obtain for the T5 spectrum has increased from $\sim 1$\% to a factor of a few.  This suggests that if we want to simplify the chemistry as much as possible while still being able to determine $J_{\rm crit}$ accurately, then the best choice is a minimal reduced network consisting of reactions 1--15, 17, 22, and 23. On the other hand, if we want to be a little more conservative, in view of the fact that the physical conditions encountered in more realistic simulations are not perfectly reproduced by our one-zone model, then we should retain all 26 of the reactions listed in Table~\ref{tab:reduce}.

\subsection{Sensitivity to the treatment of H$_{2}^{+}$ excitation}
\label{h2p_excite}
As we discuss in more detail in the Appendix, the rate at which H$_{2}^{+}$ is destroyed by processes such as dissociative recombination or photodissociation is highly sensitive to the vibrational state of the molecular ions: it is far easier to destroy H$_{2}^{+}$ ions in states with $v \gg 0$ than ones that are in the $v=0$ vibrational ground state. Therefore, any model of the chemistry of H$_{2}^{+}$ in primordial gas should also account, at least approximately, for the degree of excitation of the ions. In the chemical model presented in this paper, this is done in a simple but approximate fashion, for reasons of computational convenience. However, in principle one can construct far more complex and accurate models \citep[see e.g.][]{cop11}. An obvious question is therefore whether this complexity is necessary if one wants to accurately determine $J_{\rm crit}$, or whether the simple treatment used in this paper suffices. 

To answer this question, we performed two additional sets of runs. In the first set, we artificially set the H$_{2}^{+}$ critical density to a very large number, so that all of the chemical rates used for the ion were the ones for H$_{2}^{+}$ in its ground state. In the second set of runs, we instead set the 
H$_{2}^{+}$ critical density to zero, so that the local thermodynamic equilibrium (LTE) rates were used for all of the reactions involving H$_{2}^{+}$ as a reactant. These two limiting cases should bracket the real behaviour of the H$_{2}^{+}$ ions in the gas, and so by comparing the results of these two sets of runs, we can establish how sensitive $J_{\rm crit}$ is to the details of the excitation of the H$_{2}^{+}$ ions.

\begin{table}
\caption{Dependence of $J_{\rm crit}$ on our treatment of H$_{2}^{+}$ excitation
\label{tab:h2p_excite}}
\begin{tabular}{cccc}
\hline 
&  \multicolumn{3}{c}{$J_{\rm crit}$} \\
Run & Fiducial & $v=0$ & LTE \\ 
\hline
1 & 17.0 & 17.6 & 14.2 \\
2 & 18.0 & 18.6 & 14.5 \\
3 & 18.0 & 18.6 & 14.5 \\
4 & 1640 & 1650 & 1560 \\
5 &1630 & 1640 & 1560 \\
6 &1630 & 1640 &1560 \\
\hline
\end{tabular}
\end{table}

We show the results from these two sets of runs in Table~\ref{tab:h2p_excite}, along with the results from our fiducial model for comparison. We see that there is only a small difference between the values of $J_{\rm crit}$ that we obtain using our fiducial model and those from the $v=0$ model. This is not surprising, since the question of whether or not enough H$_{2}$ forms to cool the gas is largely determined by the chemistry occurring at densities $n < n_{\rm crit}$, where the $v=0$ rates are a good approximation. If we instead adopt LTE rates throughout, we find a larger difference between the models, approximately 20\% in the runs with the T4 spectrum and $\sim 5$\% in the runs with the T5 spectrum. As this represents a worst case, the real error introduced  is almost certainly smaller than this. Our simplified treatment of  H$_{2}^{+}$ excitation therefore does not represent a major source of error in our determination of $J_{\rm crit}$.

\subsection{Importance of dissociative tunneling}
\label{tunnel}
Our analysis in Section~\ref{reduce} showed that the collisional dissociation of H$_{2}$ has one of the largest weights in our reduced chemical network, and therefore it is very important to treat this process accurately. As discussed at some length in e.g.\ \citet{msm96}, there are two distinct processes that contribute to the total collisional dissociation rate. The first is direct collisional dissociation, where the H$_{2}$ molecule undergoes a transition from a bound state directly into the continuum of classically unbound states. The second is dissociative tunneling, where the H$_2$ molecule is excited into a quasi-bound state -- a state that has an internal energy larger than is required for dissociation, but which is separated from the continuum by a barrier in the effective potential. Quantum tunneling through this barrier then leads to spontaneous dissociation. In low density gas this process of dissociative tunneling can make a major contribution to the total collisional dissociation rate.

Some studies of the direct collapse model for the formation of massive black holes include only the effects of direct collisional dissociation, and not those of dissociative tunneling \citep[e.g.][]{sbh10}, but as \citet{latif14} point out, this potentially introduces significant uncertainty into the resulting values derived for $J_{\rm crit}$. In an effort to quantify this uncertainty, we re-ran our set of six simulations using our reduced chemical model, but with the effects of dissociative tunneling disabled. We found that in this case, $J_{\rm crit} = 33.1$ in runs 1--3 and $J_{\rm crit} = 2780$ in runs 4--6. In other words, the failure to account for this process when modelling H$_{2}$ dissociation in this scenario can introduce an error of almost a factor of two into $J_{\rm crit}$.

\subsection{Comparison with other chemical models}
It is interesting to investigate whether there are any significant differences between the reduced chemical network constructed in this paper and the other chemical networks in common usage in studies of the direct collapse model for the formation of massive black holes. The majority of existing studies use one of three treatments: the {\sc krome} astrochemistry package \citep{grassi14}, which provides a range of different networks for modelling primordial chemistry; the primordial chemistry network implemented in the {\sc enzo} adaptive mesh refinement code, described in \citet{bryan14}, which is a modified and extended version of the network outlined in \citet{abel97}; or the network introduced in \citet{om01}. We compare our reduced network to each of these treatments below.

\begin{table}
\caption{Comparison of $J_{\rm crit}$ determined using our reduced
model with that determined using several other simplified models
from the literature \label{tab:compare}}
\begin{tabular}{ccccc}
\hline 
&  \multicolumn{4}{c}{$J_{\rm crit}$} \\
Run & This work & {\sc krome} & {\sc enzo} & \citet{om01} \\ 
\hline
1 & 17.1 & 8.9  & 21.5 & 17.1 \\
2 & 18.0 & 14.7  & 26.7 & 18.3 \\
3 & 18.1 & 14.9  & 26.8 & 18.3 \\
4 & 1640 & 1120 & 1930 & 3040 \\
5 &1630 & 1160 & 1940 & 3040 \\
6 &1630 & 1160 & 1940 & 3040 \\
\hline
\end{tabular}
\end{table}

\subsubsection{The {\sc krome} package}
The {\sc krome} astrochemistry package is a Python-based pre-processing system that converts a simple textual description of an
astrochemical network into a set of subroutines for the solution of the chemical rate equations for the species contained in that 
network. In addition, {\sc krome} also contains support for a large number of different heating and cooling processes, 
allowing the thermal energy equation to be solved along with the chemical rate equations.
{\sc krome} is still undergoing active development, but in our discussion here we refer to the state of the code at
the time of writing;\footnote{Specifically, on January 3rd, 2015, at which time the most recent commit was 
{\tt 53a32c2ab5d53994fe4893676dbfbcf08125c23d}}
note that this differs in some respects from
the previous version of the package described in \citet{grassi14}.

The {\sc krome} package provides a number of example networks, including several designed for modelling primordial gas.
Unfortunately, these come with very little documentation, and so it is not immediately obvious which network is the
best choice for which application, or even which (if any) of these networks has been used in studies such as \citet{latif14}, 
\citet{vb14}, or \citet{latif15} that have been carried out using {\sc krome}.
However, some investigation allows us to immediately cut down the number of possibilities. We can clearly ignore any of
the primordial networks that do not include the photodissociation of H$^{-}$ or H$_{2}$, since these are crucial
processes in the case of the T4 or T5 spectra, respectively. This leaves us with only two networks to consider:
the {\em react\_primordial\_photoH2} network and the {\em react\_xrays} network. 
  
The  {\em react\_primordial\_photoH2} network consists of reactions 1--8, 11, 12, 15, 22, 23, 25 and 26 from our reduced 
network, along with several additional reactions, largely involving the ionization and recombination of helium and
the chemistry of D, D$^{+}$ and HD, that our results demonstrate are not important for the determination of 
$J_{\rm crit}$. The {\em react\_xrays} network consists of the reactions mentioned above, plus two more from 
our reduced network, reactions 9 and 18. It also accounts for charge transfer between hydrogen and helium,
and includes a more extensive treatment of deuterium chemistry, but once again, our results suggest that these
processes are unimportant for determining $J_{\rm crit}$.

We therefore see that compared to our reduced network, the {\em react\_xrays} network is missing reactions
10, 13--14, 16--17, 19--21 and 24, while the {\em react\_primordial\_photoH2} network is missing these plus
also reactions 9 and 18. To quantify the effect of omitting these reactions on the determination of $J_{\rm crit}$,
we have rerun our set of six simulations using only the reactions contained in the {\em react\_xrays} network,
and determined the value of $J_{\rm crit}$ in each case. Note that when doing so, we use the rate coefficients
for each reaction taken from our chemical model, which are not always the same as those adopted by the
{\sc krome} model. This is because we are interested here only in the effects of omitting some of the reactions
that we have determined are important, and not on quantifying the uncertainty introduced into $J_{\rm crit}$
by differences in our choice of rate coefficients. We examine the latter issue at some length in a companion
paper \citep{glover15}.

The results of our comparison are shown in Table~\ref{tab:compare}. We see that  the {\sc krome} network systematically 
underestimates $J_{\rm crit}$. The discrepancy is worst in run 1, where the error is almost a factor of two,
but even in the best case, the error is $\sim 20$\%. We have also carried out a similar comparison using
the set of reactions in the {\em react\_primordial\_photoH2} network, but find that in this case we recover
very similar results. 
We have investigated the source of this discrepancy and find that almost all of it is
accounted for by the omission of reactions 10 and 14 from the {\sc krome} model. If we add these two
reactions to the set included in the {\em react\_xrays} network, then we can reproduce the values of $J_{\rm crit}$
that we obtain with our minimal model to within $\sim$1\%.

\subsubsection{The {\sc enzo} network}
The primordial chemistry network implemented in the {\sc enzo} code consists of the same subset of the
reactions from our minimal model as in the {\sc krome} {\em react\_xrays} network, i.e.\ reactions
1--8, 11, 12, 15, 22, 23, 25 and 26. As in the case of the {\sc krome} networks, it also includes a number of
reactions involving the ionization and recombination of helium and the chemistry of deuterium that play
no role in the determination of $J_{\rm crit}$. The only difference between the {\sc enzo} network and
the the {\sc krome} {\em react\_xrays} network that is relevant for our present study is that the {\sc krome}
network accounts for the collisional dissociation of H$_{2}$ via dissociative tunneling as well as direct
dissociation into the continuum, whereas the {\sc enzo} model only includes the latter process. 

In Table~\ref{tab:compare}, we compare the values of $J_{\rm crit}$ that we obtain using the {\sc enzo}
model with those that we obtain using our reduced model. We find that the {\sc enzo} model systematically
overestimates $J_{\rm crit}$ by between 20\% and 50\%. Interestingly, the mean difference is smaller than 
with the {\sc krome} model, despite the fact that the {\sc enzo} model is less complete than the {\sc krome}
model. The reason for this is that while the neglect of reactions 10 and 14 in the {\sc enzo} model tends
to decrease our estimate of $J_{\rm crit}$, the neglect of dissociative tunneling has the opposite effect,
as we have seen already in Section~\ref{tunnel}. Therefore, the two effects cancel to some extent, and so the
overall difference with the results of our reduced model is less than we would initially expect.

\subsubsection{The \citet{om01} network}
The chemical network adopted by \citet{om01} in his pioneering study of the effects of a strong extragalactic
radiation field on the gravitational collapse of primordial gas includes reactions 1--10, 12, 15, 18, 22 and 26
from our reduced network. It also includes a number of additional reactions such as the three-body recombination
of H$^{+}$, or the collisional ionization of electronically excited hydrogen that are unimportant at the densities
at which the value of $J_{\rm crit}$ is set, but which play important roles in regulating the ionization state and
thermal evolution of the gas at much higher densities. The \citet{om01} network, which for the sake of brevity
we refer to hereafter as the O01 network, or an updated version of it,
has subsequently been used in a number of different studies of aspects of the direct collapse model for black
hole formation \citep[e.g.][]{om08,io11,io12,iot14,soi14}. 

In Table~\ref{tab:compare}, we compare the results we obtain for $J_{\rm crit}$ when we use the set of
reactions contained in the O01 network with the values obtained using our reduced network. 
In performing this comparison, we have assumed that the O01 network accounts for both
the direct collisional dissociation of H$_{2}$ and its destruction by dissociative tunneling. This was
not the case in the original version of the network, which used a highly outdated treatment of H$_{2}$
collisional dissociation taken from \citet{ls83}. More recent studies use an updated treatment based
on \citet{msm96}, but do not clarify whether they include only the direct collisional dissociation 
or also the dissociative tunneling term.

We see from Table~\ref{tab:compare} that there is very good agreement between the results of our
reduced network and those obtained using the O01 network in the case of the simulations
performed using the T4 spectrum. In these runs, the maximum difference between the two models
is around 2\%. The main reason that the O01 model performs much better than the {\sc krome}
or {\sc enzo} models in this case is that it includes the effects of reaction 10, which, as we have already
seen, significantly contributes to the ionization state of the gas at the densities where the value of
$J_{\rm crit}$ is set. 

In the case of the runs performed using the T5 spectrum, however, we see that there is a difference of
almost a factor of two between the results from our reduced model and those from the O01
model. This difference is driven almost entirely by the fact that the O01 model does not include
reaction 11 from our reduced model, the collisional detachment of H$^{-}$ by H:
\begin{equation}
{\rm H^{-} + H} \rightarrow {\rm H + H + e^{-}}.
\end{equation}
Although this reaction generally occurs more slowly than the associative detachment reaction responsible
for forming H$_{2}$ from H$^{-}$ (reaction number 3 in our reduced network), at the temperatures reached 
in gas with $J_{21} \sim J_{\rm crit}$, the difference between the two rates is less than an order of magnitude.
In the runs with the T4 spectrum, both reactions have rates that are small compared to the photodetachment
rate (reaction 7), and hence including reaction 11 in the model makes little difference to the equilibrium H$^{-}$
abundance or the H$_{2}$ formation rate. On the other hand, in runs performed using the T5 spectrum,
H$^{-}$ photodetachment is much less important and so reaction 11 plays a much more important role in
regulating the H$^{-}$ abundance. Its omission from the chemical network leads to one overestimating the
H$^{-}$ abundance and hence overestimating the H$_{2}$ formation rate. Consequently, a larger value of
$J_{21}$ is required in order to suppress H$_{2}$ cooling, and so the resulting values for $J_{\rm crit}$
are systematically larger. We have verified this by re-running the T5 models with a modified version of
the O01 network that includes reaction 11. We find that in this case, the difference between the value
of $J_{\rm crit}$ that we obtain for the modified O01 model and the one that we obtain for our reduced
model is only $\sim$5\%, with this remaining difference mostly likely being due to the fact that the O01
model does not include the effects of the collisional ionization of H by collisions with He (reaction 14).

\section{Summary}
\label{sum}
In this paper, we have attempted to identify the set of chemical reactions that it is essential to
include in any chemical network used to determine $J_{\rm crit}$, the critical UV flux required to
suppress H$_{2}$ cooling in atomic cooling halos with $T_{\rm vir} > 10^{4} \: {\rm K}$. To do
this, we have made use of the reaction-based reduction algorithm developed by \citet{Wiebe03}.
This has previously been applied to the study of the chemistry of the local interstellar medium, 
but to the best of our knowledge, the present paper marks its first use in the study of the chemistry
of primordial gas.

Using this reduction technique with a conservative choice of $\epsilon = 10^{-4}$ for the reaction 
weight below which we do not retain reactions in our reduced network, we find that we can reduce
our initial 30 species, $\sim 400$ reaction network to a reduced network containing only eight chemical
species linked by 26 reactions. We have verified that simulations carried out using this reduced network
predict essentially identical values of $J_{\rm crit}$ to those carried out using our full network. We have
also explored the effect of varying $\epsilon$, and find that we continue to be able to predict $J_{\rm crit}$
accurately as long as $\epsilon \leq 10^{-2}$. Setting $\epsilon = 10^{-2}$ allows us to produce an
even smaller ``minimal'' reduced network, containing the same set of eight chemical species, but now
linked by only 18 reactions. 

Most of the reactions included in our reduced networks are familiar from previous studies of the
chemistry of primordial gas. The main exception is the reaction
\begin{equation}
{\rm H + H} \rightarrow {\rm H^{+} + e^{-} + H}.
\end{equation}
This was included in the original investigation by \citet{om01} of the effect of strong UV radiation fields
on the gravitational collapse of primordial gas, but has been omitted in most subsequent studies. We
have investigated the influence of this reaction and show that omitting it leads to a 20-30\% reduction
in $J_{\rm crit}$. We also confirm the previous finding of \citet{latif14} regarding the importance of accounting 
for dissociative tunneling when treating H$_{2}$ collisional dissociation, and show that omitting this process
leads to one overestimating $J_{\rm crit}$ by almost a factor of two.

We have compared the values for $J_{\rm crit}$ that we recover using our reduced network with those
that we obtain using several other chemical networks that have previously been adopted in studies of
the direct collapse model: the {\em react\_xrays} network from the {\sc krome} astrochemistry package,
the {\sc enzo} primordial chemistry network, and the \citet{om01} network. In carrying out this comparison
we have updated the chemical rate coefficients used in these networks to match those used in the current
paper, to allow us to focus on the uncertainties introduced by differences in the set of reactions chosen
to construct the different chemical networks. 
We find that the \citet{om01} network predicts $J_{\rm crit}$ accurately for the runs with a T4 spectrum, 
but significantly overestimates it for runs with a T5 spectrum, owing primarily to its neglect of the reaction
\begin{equation}
{\rm H^{-} + H} \rightarrow {\rm H + H + e^{-}}.
\end{equation}
We also show that the {\sc enzo} network tends to predict values of $J_{\rm crit}$ that are 20--50\% larger
than those predicted by our reduced model, while the {\sc krome} {\em react\_xrays} network predicts 
values that are 20--50\% smaller. 
The total uncertainty introduced into estimates of $J_{\rm crit}$ given in the literature due to differences
between the chemical networks adopted by different studies can therefore approach a factor of three.

To put this number into context, we note that in the regime relevant for the direct collapse model,
the probability of an atomic cooling halo being exposed to a local radiation field with a strength
$J_{21} > J_{\rm crit}$ is a strongly decreasing function of $J_{\rm crit}$ \citep{dfm14}.
For example, \citet{it14} show that for $10^{3} < J_{\rm crit} < 10^{4}$, this probability scales approximately
as $P \propto J_{\rm crit}^{-5}$. A factor of three uncertainty in $J_{\rm crit}$ can therefore potentially
correspond to a factor of $\sim 200$ uncertainty in the cosmological number density of massive black
holes formed by direct collapse. 


Finally, we stress that this uncertainty can be eliminated from future studies of the direct collapse model
simply by ensuring that the chemical network adopted includes the full set of chemical reactions that are 
important for determining $J_{\rm crit}$. We therefore recommend that in future work, researchers take
care to include, at the very least, all of the reactions  making up the minimal reduced model described
in Section~\ref{epsilon}.

\section*{Acknowledgements}
The author would like to thank T.~Hartwig, S.~Khochfar, R.~Klessen, and M.~Volonteri for useful conversations regarding the
physics of black hole formation in the high-redshift Universe. Additional thanks also go to T.~Hartwig for his comments on an
earlier draft of this paper. Special thanks go to B.~Agarwal and B.~Smith for several 
discussions regarding the possible impact of chemical uncertainties on $J_{\rm crit}$ that inspired the author to carry out the
work described here. Financial support for this work was provided by the Deutsche Forschungsgemeinschaft  via
SFB 881, ``The Milky Way System'' (sub-projects B1, B2 and B8) and SPP 1573, ``Physics of the Interstellar Medium'' (grant number GL 668/2-1), 
and by the European Research Council under the European Community's Seventh Framework Programme (FP7/2007-2013) via the 
ERC Advanced Grant STARLIGHT (project number 339177).

\appendix
\section{Revisions to the thermal model}
\label{app:cool}

We have improved on the thermal model introduced in \citet{ga08}
and \citet{Glover09} by updating the rates of two of the cooling processes in 
an effort to use the most up-to-date data available. Details of our changes
are given below.

\subsection{Updated cooling rates}
\subsubsection{Collisional excitation of H$_{2}$ by electrons}
The cooling rate due to collisions between H$_{2}$ molecules and free electrons
that we used in \citet{Glover09} was taken from \citet{ga08} and was based on
rather old data from  \citet{drd83}. We have now updated our treatment of this process 
and use a rate based on data from the recent compilation of \citet{yoon08}. The resulting 
cooling rate  is well fit by the expression
\begin{eqnarray}
\log \Lambda_{\rm H_{2}, {\rm e}} & = & \mbox{} -21.928796 + 16.815730 \log(T_{3}) \nonumber \\
& & \mbox{} + 96.743155 \log(T_{3})^{2}  \nonumber \\
& & \mbox{} + 343.19180 \log(T_{3})^{3} \nonumber \\
& & \mbox{} + 734.71651 \log(T_{3})^{4}  \nonumber \\
& & \mbox{} + 983.67576 \log(T_{3})^{5} \nonumber \\
& & \mbox{} + 801.81247 \log(T_{3})^{6}  \nonumber \\
& & \mbox{} + 364.14446 \log(T_{3})^{7} \nonumber \\
& & \mbox{} + 70.609154 \log(T_{3})^{8},
\end{eqnarray}
at temperatures $100 < T < 500 \: {\rm K}$, and by 
\begin{eqnarray}
\log \Lambda_{\rm H_{2}, {\rm e}} & = & \mbox{} -22.921189 + 1.6802758 \log(T_{3}) \nonumber \\
& & \mbox{} + 0.93310622  \log(T_{3})^{2} \nonumber \\
& & \mbox{} + 4.0406627  \log(T_{3})^{3} \nonumber \\
& & \mbox{}  - 4.7274036  \log(T_{3})^{4} \nonumber \\
& & \mbox{} - 8.8077017  \log(T_{3})^{5} \nonumber \\
& & \mbox{}  + 8.9167183  \log(T_{3})^{6} \nonumber \\
& & \mbox{} + 6.4380698  \log(T_{3})^{7} \nonumber \\
& & \mbox{}  - 6.3701156  \log(T_{3})^{8},
\end{eqnarray}
at temperatures $500 < T < 10^{4} \: {\rm K}$, where $T_{3} = T / 1000 \: {\rm K}$.
These fits are accurate to within 5\% over the quoted temperature range. This
revised treatment yields less cooling at all temperatures than the rate given in
\citet{ga08}, with the effect being particularly pronounced at temperatures around
1000~K. However, the effect on the total H$_{2}$ cooling rate is less significant,
since H$_{2}$--H$^{+}$ collisions lead to substantially more cooling than 
H$_{2}$--e$^{-}$ collisions in gas where $n_{\rm H^{+}} \simeq n_{\rm e}$.

\subsubsection{Collisional excitation of H$_{2}$ by protons}
We have also updated our treatment of the cooling rate due to collisions between 
H$_{2}$ molecules and protons. The treatment in \citet{Glover09} was based on
\citet{ga08} and made use of data from \citet{ger90} and \citet{kr02}. Our new treatment
makes use of the excitation rates recently calculated by \citet{hon11,hon12} for the transitions 
for which these are available, supplementing them with data from \citet{ger90} and \citet{kr02} 
for those transitions for which newer data is not available. The resulting cooling rate is
well fit by the expression
\begin{eqnarray}
\log \Lambda_{\rm H_{2}, H^{+}} & = & -22.089523 + 1.5714711 \log(T_{3}) \nonumber \\
& & \mbox{} + 0.015391166 \log(T_{3})^{2} \nonumber \\
& & \mbox{} - 0.23619985 \log(T_{3})^{3} \nonumber \\
& & \mbox{}  - 0.51002221 \log(T_{3})^{4} \nonumber \\
& & \mbox{} + 0.32168730 \log(T_{3})^{5},
\end{eqnarray}
for temperatures in the range $10 < T < 10^{4} \: {\rm K}$. Again, the revised treatment
yields less cooling at all temperatures than the rate given in \citet{ga08}. In conditions 
where H$_{2}$--H$^{+}$ collisions make the dominant contribution to the H$_{2}$
cooling function, the total H$_{2}$ cooling rate can be as much as a factor of two
smaller.

\section{Revisions to the chemical model}
\label{app:chem}
We have improved on the chemical model introduced in \citet{Glover09}
by including three new reactions and updating the reaction rates used for
a further eight reactions. Details of our changes are given below.

\subsection{New reactions}

\subsubsection{Collisional ionization of atomic hydrogen}
Because the rate at which electrons bring about the collisional ionization of atomic hydrogen
\begin{equation}
{\rm H + e^{-}} \rightarrow {\rm H^{+} + e^{-} + e^{-}}
\end{equation}
is substantially faster than the ionization rate due to H-H collisions
\begin{equation}
{\rm H + H} \rightarrow {\rm H^{+} + e^{-} + H},
\end{equation}
the latter process has been omitted from most chemical models of primordial gas. 
However, it can actually dominate if the fractional ionization of the gas falls below 
$x \sim 10^{-4}$, or if there is a substantial population of H atoms in excited electronic 
states \citep{om01}. We therefore include this process in our revised chemical
model, using the following rate coefficient \citep{lcs91,om01}.
\begin{equation}
k_{\rm ci, H} = 1.2 \times 10^{-17} T^{1.2} \exp \left(\frac{-157800}{T} \right) \: {\rm cm^{3} \: s^{-1}}.
\end{equation}
This rate coefficient is based on the experimental cross-sections measured by 
\citet{gvz87}, and assumes that both hydrogen atoms are in their ground state.
As we discuss in more detail in Paper II, a number of other versions of
the rate coefficient for this reaction can be found in the literature \citep{drawin69,hm79,hm89,ks91,soon92,barklem07}.
These expressions differ by large amounts at the temperatures relevant for this
study and it is unclear which expression gives the best description of the actual 
behaviour of this reaction. In this paper, we have chosen to use the version of
the rate coefficient given in \citet{lcs91} for consistency with the earlier study of
\citet{om01}, but in Paper II we explore the sensitivity of $J_{\rm crit}$ to this choice.

We also include the collisional ionization of atomic hydrogen by collisions with
neutral helium
\begin{equation}
{\rm H + He} \rightarrow {\rm H^{+}  + e^{-} + He}.
\end{equation}
We use the rate coefficient computed by \citet{lcs91} using cross-sections from
\citet{vla81}:
\begin{equation}
k_{\rm ci, He} = 1.75 \times 10^{-17} T^{1.3} \exp \left(\frac{-157800}{T} \right) \: {\rm cm^{3} \: s^{-1}}.
\end{equation}
Again, this expression assumes that all of the hydrogen atoms are in their ground
state.

In practice, trapping of Lyman-$\alpha$ photons in the collapsing gas owing to
the high optical depth in the Lyman-$\alpha$ line leads to the existence of a
non-zero population of hydrogen atoms in excited electronic states \citep[see
e.g.][]{om01,ssg10}. However, at the densities of interest in this study, only a
very small fraction (of order $10^{-10}$ or less) of the total number of hydrogen
atoms are found in these states \citep{ssg10}, and so there is no need to account
in the chemical model for the effects of collisional ionization out of these states. 

\subsubsection{Collisional dissociation of H$_{2}^{+}$ by atomic hydrogen}
\label{h2p_cd}
We now account for the destruction of H$_{2}^{+}$ by the reaction
\begin{equation}
{\rm H_{2}^{+} + H} \rightarrow {\rm H^{+} + H + H}.
\end{equation}
The rate at which this reaction proceeds depends on the internal excitation of
the H$_{2}^{+}$ molecular ion. At low densities, it is safe to assume that all of
the ions are in the vibrational ground state. In this limit, we adopt the following
rate coefficient for this reaction
\begin{equation}
k_{\rm cd, 0} = 1.54 \times 10^{-12} T^{0.45} \exp \left(\frac{-31000}{T} \right) \: {\rm cm^{3} \: s^{-1}},
\end{equation}
which we have derived based on the cross-sections presented in \citet{kj03}.
At high densities, on the other hand, the vibrational level populations of the
H$_{2}^{+}$ ion approach their local thermodynamic equilibrium (LTE) values.
In this limit, we use the LTE rate for this reaction derived by \citet{cop11},
which was again based on the \citet{kj03} cross-sections. At intermediate
densities, we interpolate between these two limiting cases by adopting a
rate coefficient of the form
\begin{equation}
k_{\rm cd} = k_{\rm cd, LTE} \left(\frac{k_{\rm cd, 0}}{k_{\rm cd, LTE}} \right)^{\alpha},
\end{equation}
where $k_{\rm cd, LTE}$ is the rate coefficient in the LTE limit, 
$\alpha = (1 + n / n_{\rm crit})^{-1}$,  and $n_{\rm crit}$ is the critical density for  
H$_{2}^{+}$ (i.e.\ the density at which the effects of collisional
de-excitation and radiative de-excitation become comparable).
 
As explained in \citet{Glover09}, in primordial gas the dominant
contributions to the collisional excitation or de-excitation of H$_{2}^{+}$
come from collisions with atomic hydrogen or with electrons. When 
collisions with H atoms dominate, we can estimate the critical density
by taking the ratio of the cooling rates due to H$_{2}^{+}$ in the LTE
and low density limits, using the expressions for these given in 
\citet{Glover09}. The resulting values are reasonably well approximated
by the expression
\begin{equation}
n_{\rm crit, H} \simeq 400 T_{4}^{-1} \: {\rm cm^{-3}},
\end{equation}
where $T_{4} = T / 10^{4} \: {\rm K}$. A similar procedure applied to collisions 
with electrons yields a critical density of electrons that does not vary significantly
with temperature, and that is roughly an order of magnitude smaller than 
$n_{\rm crit, H}$,
\begin{equation}
n_{\rm crit, e} \simeq 50 \: {\rm cm^{-3}}.
\end{equation}
We combine the contributions from atomic hydrogen and free electrons by 
taking a weighted harmonic average, yielding
\begin{equation}
n_{\rm crit} = \left(\frac{x_{\rm H}}{n_{\rm crit, H}} + \frac{x_{\rm e}}{n_{\rm crit, e}} \right)^{-1}.
\end{equation}

Although this is a somewhat crude approximation, it is sufficient for our purposes, since
$J_{\rm crit}$ is not very sensitive to the treatment adopted for H$_{2}^{+}$ excitation
(see Section~\ref{h2p_excite}).

\subsection{Updated reaction rates}
\label{updated}
Since the publication of \citet{Glover09}, new experimental and/or theoretical data has 
become available for a number of different reactions. We have therefore updated the
rates adopted for several of the included reactions, as detailed below.

\subsubsection{Associative detachment of H$^{-}$ with H}
The rate coefficient for the reaction 
\begin{equation}
{\rm H^{-} + H} \rightarrow {\rm H_{2} + e^{-}}
\end{equation}
was recently measured by \citet{kreck10} for temperatures in the range
$1 < T < 10^{4}$~K using a merged-beam approach
\citep[see also][]{bru10,miller11}. The estimated systematic error in their
measurements is around 25\%. 
Their results disagree with the flowing afterglow results
of \citet{martinez09} at 300~K by a factor of two, for reasons which remain 
uncertain, but agree well with the measurements made by \citet{ger12}
at temperatures $10 < T < 150$~K using an ion trap. We therefore adopt
the rate coefficient of \citet{kreck10} for this reaction. We also adopt the same
rate coefficients for the isotopic variants of this reaction (i.e.\ reactions
where one or both H atoms are replaced by D atoms), following
\citet{miller12}, who found that there is no significant isotope 
effect in the cross-section for this reaction.

\subsubsection{Mutual neutralization of H$^{-}$ with H$^{+}$}
Until relatively recently, the rate of the mutual neutralization reaction
\begin{equation}
{\rm H^{-} + H^{+}} \rightarrow {\rm H + H}
\end{equation}
at low temperatures was quite unclear. \citet{gsj06} surveyed the
range of rates given for this reaction in the astrophysical literature
as of 2006, and showed that there was almost an order of magnitude
scatter in the values. Fortunately, the last few years have seen a
significant improvement in this area. New theoretical \citep{sle09} and 
experimental \citep{urb12} determinations of the rate coefficient yield values that are in good agreement 
with the ones derived by \citet{cdg99} from the cross-sections of \citet{fk86}. We 
therefore adopt the \citet{cdg99} rate coefficient for use in our study.

\subsubsection{Dissociative recombination of H$_{2}^{+}$}
The rate coefficient used in \citet{Glover09} for the reaction
\begin{equation}
{\rm H_{2}^{+} + e^{-}} \rightarrow {\rm H + H}
\end{equation}
was a fit made by \citet{abel97} to the data of \citet{schn94,schn97}.
It assumes that the H$_{2}^{+}$ molecular ions are in their vibrational
ground state. However, this is true only at low densities. At high
densities, the H$_{2}^{+}$ level populations tend towards their
local thermodynamic equilibrium (LTE) values, and the resulting
dissociative recombination rate can be significantly larger 
\citep[see e.g.\ the comparison of low density and LTE rates in
Figure 9 of][]{cop11}. To account for this, we use a density
dependent rate coefficient of the form
\begin{equation}
k_{\rm dr} = k_{\rm dr, LTE} \left(\frac{k_{\rm dr, 0}}{k_{\rm dr, LTE}} \right)^{\alpha},
\end{equation}
where $k_{\rm dr, 0}$ is the rate coefficient in the low density limit (taken
as before from \citealt{abel97}), $k_{\rm dr, LTE}$ is the rate coefficient
in the LTE limit (taken from \citealt{cop11}, based on data from
\citealt{takagi02}), $\alpha = (1 + n / n_{\rm crit})^{-1}$, 
and $n_{\rm crit}$ is the critical density for  H$_{2}^{+}$,
calculated as outlined in Section~\ref{h2p_cd} above.
 
\subsubsection{Collisional dissociation of H$_{2}$ by H}
We now use the fitting formulae given in \citet{msm96} to determine
the rate coefficient for the reaction
\begin{equation}
{\rm H_{2} + H} \rightarrow {\rm H + H + H},
\end{equation}
in place of the combination of data from several sources used in \citet{Glover09}.
Importantly, we include the contribution to the total H$_{2}$ dissociation rate due
to excitation to a quasi-bound state followed by dissociative tunneling to an
unbound state ($\gamma_{\rm dt}$ in the notation of \citealt{msm96}). At low
temperatures ($T < 4500 \: {\rm K}$ in the low density limit), this effect is more
important than direct collisional dissociation and 
it is important to account for it if one wants to determine $J_{\rm crit}$ accurately
(see \citealt{latif14} or Section~\ref{tunnel} of the present paper).

\subsubsection{Charge transfer from H$^{+}$ to H$_{2}$}
\citet{Glover09} adopted a rate coefficient for the reaction
\begin{equation}
{\rm H_{2} + H^{+}} \rightarrow {\rm H_{2}^{+} + H}
\end{equation}
that was taken from \citet{savin04a,savin04b} and that assumes that all of the H$_{2}$
molecules are in their vibrational ground state. However, in the LTE limit, the
actual rate can be as much as an order of magnitude larger \citep{cop11}.
In the present study, we therefore adopt a density-dependent rate coefficient
of the form
\begin{equation}
k_{\rm ct} = k_{\rm ct, LTE} \left(\frac{k_{\rm ct, 0}}{k_{\rm ct, LTE}} \right)^{\alpha},
\end{equation}
where $k_{\rm ct, 0}$ is the rate coefficient in the low density limit (taken from
\citealt{savin04a,savin04b}), $k_{\rm ct, LTE}$ is the rate coefficient in the LTE limit
(taken from \citealt{cop11}), $\alpha = (1 + n / n_{\rm crit})^{-1}$, 
and $n_{\rm crit}$ is the H$_2$ critical density. We conservatively assume that
the latter has the same value here as for collisions between H$_{2}$ and H.

\subsubsection{Three-body formation of H$_{2}$}
A large variety of different rate coefficients have been given in the astrophysical 
literature for the reaction
\begin{equation}
{\rm H + H + H} \rightarrow {\rm H_{2} + H},
\end{equation}
as summarized in \citet{glo08} and \citet{turk11}. Typically, these values were 
derived by taking the experimentally-measured rate for the inverse process
(collisional dissociation of H$_{2}$ by H) and then applying the principle of detailed
balance. In general, the resulting rate coefficients agree fairly well at high temperatures
(where the collisional dissociation rate can be easily measured), but disagree substantially
at low temperatures, owing in large part to the errors introduced by extrapolating the collisional 
dissociation rates outside of the measured temperature range. 

This unsatisfactory state of affairs was recently addressed by \citet{forrey13a,forrey13b}.
He computed a rate coefficient for this reaction using a new technique that does not rely
on detailed balance and that hence should be far more reliable at low gas temperatures.
The resulting rate coefficient is well fit by the simple expression \citep{forrey13a}
\begin{equation}
k_{\rm 3b} = 6 \times 10^{-32} T^{-1/4} + 2 \times 10^{-31} T^{-1/2} \: {\rm cm^{6} \: s^{-1}},
\end{equation}
and we use this value in all of our calculations. 

\subsubsection{Photodissociation of H$_{2}^{+}$}
The rate adopted by \citet{Glover09} for the process 
\begin{equation}
{\rm H_{2}^{+} + \gamma} \rightarrow {\rm H^{+} + H}
\end{equation}
assumed at T5 spectrum and also that
all of the H$_{2}^{+}$ molecular ions were in their vibrational ground state. The latter assumption
is reasonable at the redshifts of interest in this study provided that the gas density is significantly
lower than the H$_{2}^{+}$ critical density. However, as $n \rightarrow n_{\rm crit}$, it becomes
important to account for the effects of vibrational excitation, as this can potentially have a large
influence on the photodissociation rate \citep{gp98}. 

In our present study, we have therefore calculated H$_{2}^{+}$ photodissociation rates for the
limiting cases where all of the molecular ions are in the $v=0$ state and where the vibrational
level populations have their LTE values. In the $v=0$ case, we use the expression
\begin{equation}
k_{\rm pd, H_{2}^{+}, v=0} = 2.0 \times 10^{1} T_{\rm rad}^{1.59} \exp \left(-\frac{82000}{T_{\rm rad}} \right) \: {\rm s^{-1}}
\end{equation}
given by \citet{gp98}, based on data from \citet{dunn68}, to compute the photodissociation rate
for a black-body spectrum and then rescale both the strength of the spectrum and the size of the
rate by the same amount so that the resulting specific intensity at the Lyman limit is given
by $10^{-21} \: {\rm erg s^{-1} cm^{-2} Hz^{-1} sr^{-1}}$. For a T4 spectrum, this procedure yields
a photodissociation rate given by
\begin{equation}
k_{\rm pd, H_{2}^{+}, v=0} = 1.74 \times 10^{-10} J_{21} \: {\rm s^{-1}}, \label{h2p_t4_v0}
\end{equation}
while for a T5 spectrum we have
\begin{equation}
k_{\rm pd, H_{2}^{+}, v=0} = 5.77 \times 10^{-12} J_{21} \: {\rm s^{-1}}. \label{h2p_t5_v0}
\end{equation}
In the LTE case, the H$_{2}^{+}$ photodissociation rate depends on the gas temperature, which
we cannot assume is the same as the radiation temperature. We therefore cannot use the
formula given in \citet{gp98}, which does assume that $T_{\rm gas} = T_{\rm rad}$, but instead
compute the required rates using the cross-sections given in \citet{stancil94}. These are valid
for temperatures in the range $3150 < T < 25200 \: {\rm K}$. We find that for temperatures in
this range, the H$_{2}^{+}$ photodissociation rate for a T4 spectrum is well-fit in the temperature
range $3150 < T < 9000 \: {\rm K}$ by the following expression
\begin{equation}
k_{\rm pd, H_{2}^{+}, LTE} = 3.45 \times 10^{-8} \exp \left(-\frac{8500}{T} \right) J_{21} \: {\rm s^{-1}},
\label{h2p_t4_lte_low}
\end{equation}
and at temperatures $9000 < T < 25200 \: {\rm K}$ by the expression
\begin{equation}
k_{\rm pd, H_{2}^{+}, LTE} = 2.5 \times 10^{-7} T^{-0.22}  \exp \left(-\frac{8500}{T} \right) J_{21} \: {\rm s^{-1}}.
\label{h2p_t4_lte_high}
\end{equation}
Within the range of temperatures quoted above, the fitting error of these expressions is no more
than 10\%. At temperatures $T < 3150 \: {\rm K}$, we extrapolate using Equation~\ref{h2p_t4_lte_low} until we reach
the rate given by Equation~\ref{h2p_t4_v0}, since the $v=0$ rate is also the appropriate low $T$ limit of the LTE
rate. At temperatures $T > 25200 \: {\rm K}$, we could in principle again extrapolate using Equation~\ref{h2p_t4_lte_high},
but in practice, the gas in our models never reaches these temperatures. 

For the T5 spectrum, we use a similar procedure. In this case, the LTE rate is well-fit over the whole
temperature range $3150 < T < 25200 \: {\rm K}$ by the expression
\begin{equation}
k_{\rm pd, H_{2}^{+}, LTE} = 2.2 \times 10^{-11} T^{-0.22} J_{21} \: {\rm s^{-1}}.
\end{equation}
As before, at temperatures $T < 3150$~K, we extrapolate the rate coefficient using the same expression until
we reach the value given by Equation~\ref{h2p_t5_v0}.

Finally, in our numerical models we smoothly interpolate between the two limiting cases using 
the following expression:
\begin{equation}
k_{\rm pd, H_{2}^{+}} = k_{\rm pd, H_{2}^{+} LTE} \left(\frac{k_{\rm pd, H_{2}^{+},  v=0}}{k_{\rm pd, H_{2}^{+}, LTE}} \right)^{\alpha},
\end{equation}
where $\alpha = (1 + n / n_{\rm crit})^{-1}$,  and $n_{\rm crit}$ is the critical density of H$_2^{+}$, computed as described in
Section~\ref{h2p_cd} above.

\subsubsection{Formation of H$_{2}^{+}$ by radiative association}
Rate coefficients for the reaction
\begin{equation}
{\rm H + H^{+}} \rightarrow {\rm H_{2}^{+}} + \gamma
\end{equation}
have been computed by both \citet{rp76} and \citet{sbd93}, and agree to within 3\%. However, the analytical fits to 
this data used in most current models of primordial chemistry are rather less accurate. Many models use the fit
introduced by \citet{sk87}, which has the form
\begin{equation}
k_{\rm ra, H_{2}^{+}} = 1.85 \times 10^{-23} T^{1.8} \: {\rm cm^{3} \, s^{-1}}
\end{equation}
for $T < 6700$~K and
\begin{equation}
k_{\rm ra, H_{2}^{+}} = 5.81 \times 10^{-16} \left(\frac{T}{56200} \right)^{\eta(T)} \: {\rm cm^{3} \, s^{-1}}
\end{equation}
for $T > 6700$~K, where $\eta(T) = -0.6657 \log (T / 56200)$.
On the other hand, the \citet{Glover09} model uses instead the fit given in \citet{gp98}
\begin{eqnarray}
k_{\rm ra, H_{2}^{+}} & = & {\rm dex} \left[-19.38 - 1.523 \log T + 1.118 (\log T)^{2} \right. \nonumber \\
& & \left. - 0.1269 (\log T)^{3} \right]  \: {\rm cm^{3} \, s^{-1}}.
\end{eqnarray}
Although both of these analytical fits are based on the \citet{rp76} data, they differ from each other, and from the rates
tabulated in by \citeauthor{rp76}, by as much as 30\%. For this reason, in our current study we do not use either of these
prescriptions. Instead, we use the analytical fit given in the recent paper by \citet{latif15}, and based on the work of \citet{cop11}
\begin{eqnarray}
k_{\rm ra, H_{2}^{+}} & = & {\rm dex} \left[-18.20 - 3.194 \log T + 1.786 (\log T)^{2} \right. \nonumber \\
& & \left. - 0.2072 (\log T)^{3} \right]  \: {\rm cm^{3} \, s^{-1}}.
\end{eqnarray}
The fit gives values for the rate coefficients that agree with those tabulated by \citet{rp76} to within around 5-6\%. Moreover,
at the temperatures most relevant in this study, $T \sim 4000$--8000~K, the agreement is even better, with the fit differing from
the tabulated values by no more than 1--2\%.

\end{document}